\documentclass{svproc}
\usepackage{url}

\usepackage{subcaption}
\usepackage[numbers,sort&compress]{natbib}
\usepackage{graphicx}
\usepackage{hyperref}

\AtBeginDocument{%
  }

\begin{document}
\mainmatter
\title{Simulating hashtag dynamics with networked groups of generative agents}

\author{Abha Jha\inst{1} \and J. Hunter Priniski\inst{2} \and Carolyn Steinle\inst{2}
\and
Fred Morstatter\inst{1}}
\authorrunning{Jha et al.} 
%
\tocauthor{Abha Jha, J. Hunter Priniski, Carolyn Steinle, Fred Morstatter}
\institute{USC Information Sciences Institute, Marina del Rey CA 90292, USA,\\
\email{\{abhajha, fredmors\}@isi.edu} \\
\and University of California, Los Angeles, CA 90095, USA, \\
\email{\{priniski, csteinle\}@ucla.edu} \\
}





\maketitle

\begin{abstract}
Networked environments shape how information embedded in narratives influences individual and group beliefs and behavior. This raises key questions about how group communication around narrative media impacts belief formation and how such mechanisms contribute to the emergence of consensus or polarization. Language data from generative agents offer insight into how naturalistic forms of narrative interactions (such as hashtag generation) evolve in response to social rewards within networked communication settings. To investigate this, we developed an agent-based modeling and simulation framework composed of networks of interacting Large Language Model (LLM) agents. We benchmarked the simulations of four state-of-the-art LLMs against human group behaviors observed in a prior network experiment (Study 1) and against naturally occurring hashtags from Twitter (Study 2). Quantitative metrics of network coherence (e.g., entropy of a group's responses) reveal that while LLMs can approximate human-like coherence in sanitized domains (Study 1's experimental data), effective integration of background knowledge and social context in more complex or politically sensitive narratives likely requires careful and structured prompting. 
\end{abstract}

\section{Introduction}

Narrative interaction in networked environments shapes individual and group behaviors, both online \cite{priniski2021mapping,priniski2021rise,papacharissi_affective_2015,adams2022knowledge,wong2022cognitive} and offline \cite{kross2021social,mooijman2018moralization,priniski_darkening_2022}. A promising approach to studying these dynamics involves computational models that simulate how individuals integrate narrative information with input from their social network. Recent advances in large language models (LLMs) have sparked interest in using them as as models of human cognition and behavior \cite{abdurahman2025primer,inducing_anxiety_in_llms,llms_human_like_reasoning,priniski_pipeline_2023}. This work includes treating LLMs as models of individual cognition \cite{turning_llms_into_cognitive_models} and applying cognitive psychology methodologies, such as controlled experimental designs, to test their generative and inferential capacities \cite{cog_psychology_gpt,webb2023emergent}.

A dovetailed line of research explores LLMs as models of group behavior \cite{opinion_dynamics_using_llms,wisdom_of_partisan_crowds,network_dynamics_crowd_wisdom}. Traditionally, group cognition has been modeled using agent-based network simulations, where each network node represents an individual agent implementing explicit computational social learning mechanisms to manipulate representations and generate responses. Researchers can now extend, or even replace, computationally simple (though explanatorily powerful) agents with LLM-based ones. The first wave of modeling research on LLM-based group dynamics focuses on evaluating whether networks of communicating LLMs can reproduce well-known behavioral trends observed in human social networks. For example, networks of communicating LLMs have been shown to produce social connectivity patterns consistent with preferential attachment \cite{chang2024llms,de2023emergence,papachristou2024network}. These networks can also exhibit well-known social phenomena such as homophily; for instance, interacting AI chatbots tend to form distinct communities based on shared language features and preferences \cite{he2024artificial}. However, the magnitude and robustness of emergent group phenomena remain uncertain\cite{chang2024llms}. 

LLMs are highly expressive, and human group dynamics are deeply complex. To evaluate when and how LLM-based group behavior aligns with human patterns, it is essential to benchmark against empirical human data and to clearly specify the context in which agents operate, i.e, their training histories, prompting strategies, and communication procedures. In this article, we compare agent outputs produced by four LLM models with data from controlled human network experiments and from naturally occurring online interactions on Twitter.

\section{Methodology}

\subsection{Hashtag matching game}

Coordination games are a widely used experimental design for studying networked group dynamics in humans. For example, Centola and Baronchelli (2015) introduced the \textit{Name Game}, in which participants are shown an image of a woman’s face and rewarded for producing matching name responses with their network neighbors \cite{centola_spontaneous_2015}. This simple interaction task was designed to demonstrate how variations in network connectivity (e.g., fully connected vs. ring-like structures) influence the emergence of shared responses within a networked group. Over the course of forty interaction trials, fully connected groups consistently converged on a shared name, while ring-like groups did not.

\begin{figure}
    \centering
    \includegraphics[width=0.8\linewidth]{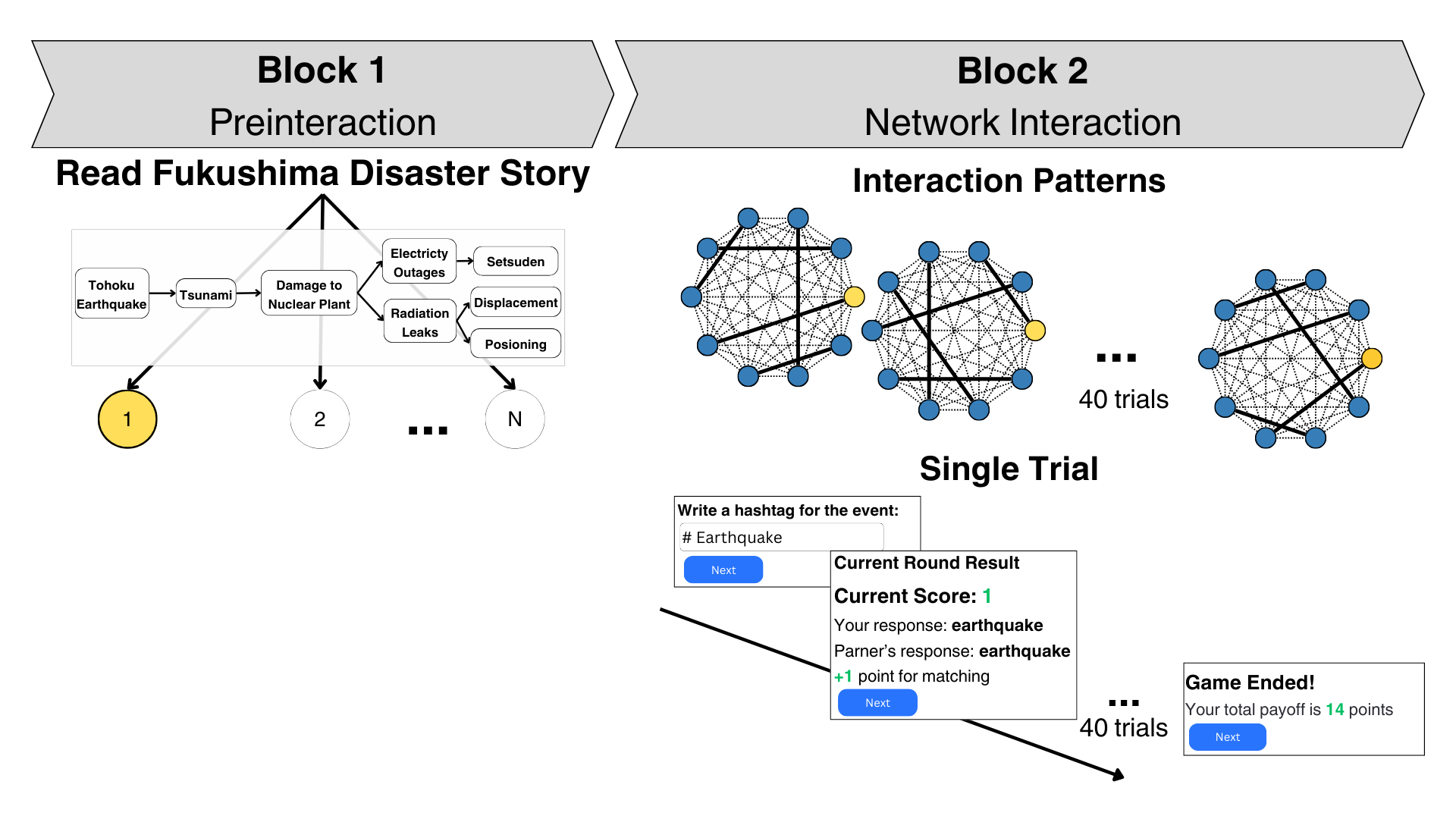}
    \caption{Procedure of the hashtag game using the Fukushima nuclear disaster narrative (Study 1) as the focal narrative and a fully-connected network group. The Hashtag Game follows a two block design. In the first block, all members of a group read a shared narrative about an event termed a \textit{focal narrative}. Note participants do not see the causal model structure presented here, but rather read a text-based narrative describing that causal structure. In the second block, participants engage in networked interaction where they are incentivized to produce matching hashtags characterizing the narrative with network neighbors.}
    \label{fig:hashtag-game-proc}
    \vspace{-1em}
\end{figure}

More recently, Priniski et al. (2024) extended the Name Game paradigm to better capture narrative interaction behaviors observed in real-world networks \cite{priniski2024online}. As shown in Figure \ref{fig:hashtag-game-proc}, the \textit{Hashtag Game} presents participants with a shared account of an event—termed the \textit{focal narrative}, and incentivizes them, across several interaction trials, to generate matching hashtags with their network neighbors. The added narrative complexity requires participants to integrate background causal knowledge with social context in order to align on shared hashtags. Hashtags were selected as the interaction behavior because they encode causal relationships and mark messages with broad semantic categories, providing a useful window into the reasoning and linguistic processes that drive narrative coordination in social networks.

\subsection{LLM-based agents}
Our simulation studies are based on the \textit{Hashtag Matching} game which allows us to evaluate how generative agent-based models integrate background knowledge (corpus statistics representations) with social context (neighbors' generated outputs) to produce shared behaviors. Specifically, we simulate networked group behavior using four different language models: DeepSeek-R1 \cite{deepseek_r1_2025}, LLaMA-3 \cite{meta_llama_3.1_8b_instruct}, Qwen2 \cite{qwen2-7b} and Gemma \cite{gemma2_2024}. These models were selected because they differ in training objectives, architectures, and training data, offering a diverse testbed for modeling group coordination. 
LLaMA-3 uses grouped-query attention (GQA), a mechanism that reduces computation by sharing key and value projections across attention heads, enabling faster inference and greater scalability and was trained on a multilingual, filtered web corpus; DeepSeek-R1 emphasizes step-by-step reasoning through chain-of-thought alignment \cite{chain_of_thought}, which involves fine-tuning on multi-step rationales, encouraging the model to generate intermediate reasoning steps that improve factual accuracy and transparency. is the only model in our study trained on large-scale bilingual (Chinese-English) data, enabling stronger cross-lingual generalization and flexibility in handling multilingual discourse. Gemma, developed by Google, was trained using reinforcement learning from human feedback (RLHF) and instruction tuning, which improves output helpfulness and alignment by optimizing model behavior to match human-preferred responses.

\subsubsection*{\textbf{Interaction Structure and Agent View}}

In our simulation studies, we prompted agents to generate hashtags in response to the focal narrative relevant to their networked interaction (Study 1: Fukushima nuclear disaster; Study 2: Philippine Presidential Election). Following the design of the \textit{Hashtag Game}, agents produced hashtags across multiple rounds of interaction. On each trial, agents were randomly paired with a partner based on the underlying network structure (described in more detail below) and re-prompted with updated social context information, i.e., the hashtags previously generated by their neighbors.

\subsubsection*{\textbf{Prompt Construction and Dynamics Across Rounds}}

Prompts were dynamically updated over the course of the interaction to integrate the focal narrative and the evolving social context, i.e., the set of hashtags generated by each agent’s partners across prior trials. For rounds $t > 1$, the prompt issued to each LLM-controlled agent incorporated the focal narrative and social context of previous interactions. (On the first round, the prompt was identical except for the absence of prior interaction content.) The prompt template was as follows:
\begin{quote}
In this experiment, you are awarded 1 point if you guess the same hashtag as your randomly assigned neighbor, and 0 points if you do not. Your goal is to earn as many points as possible.

You are in round [t] of the experiment. Your guesses and your neighbor's guesses have been as follows, represented in the CSV below:

[Interaction Table from Round t-1]

Based on this information and the event provided in round 1:

[Event Description]

Please guess a short (max 5 words) hashtag for this event. Try to match your neighbor while staying relevant to the event. You may reuse your previous hashtag, but don't always do so—especially if you believe your next neighbor might choose something different.
\end{quote}
The [Event Description] is updated based on the focal narrative for the simulation study. In study 1, the event description is the Fukushima nuclear disaster narrative; in study 2, it is the Philippine election narrative. The narrative materials constituting each study's event description, including more information on their narrative structure and content, are provide in each study's section below.






\begin{figure}[t]
    \centering
    \begin{subfigure}[t]{0.48\textwidth}
        \centering
        \includegraphics[width=\textwidth]{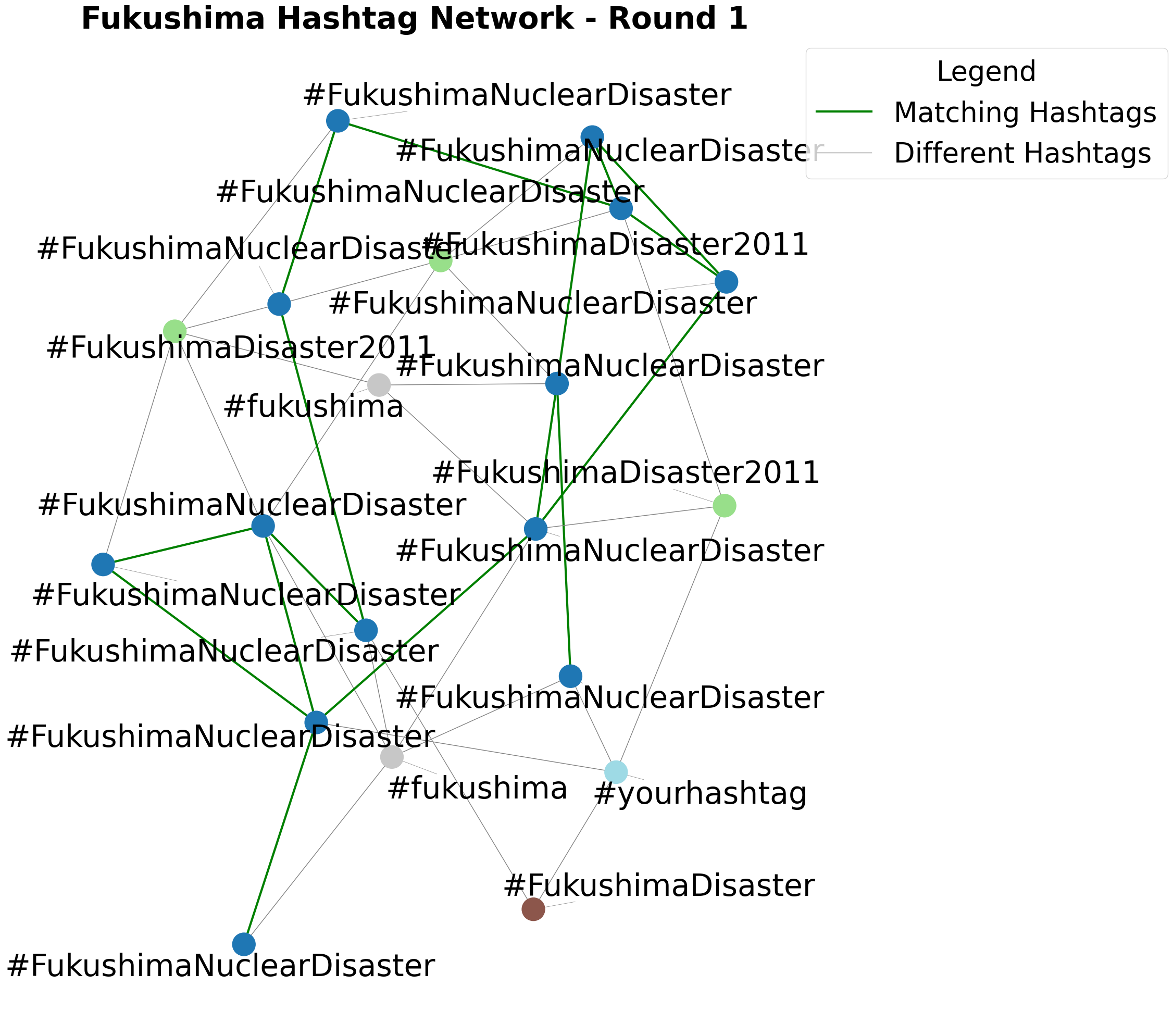}
        \caption{LLaMA Network — Round 1}
        \label{fig:network_structure_llama_round_1}
    \end{subfigure}
    \hfill
    \begin{subfigure}[t]{0.48\textwidth}
        \centering
        \includegraphics[width=\textwidth]{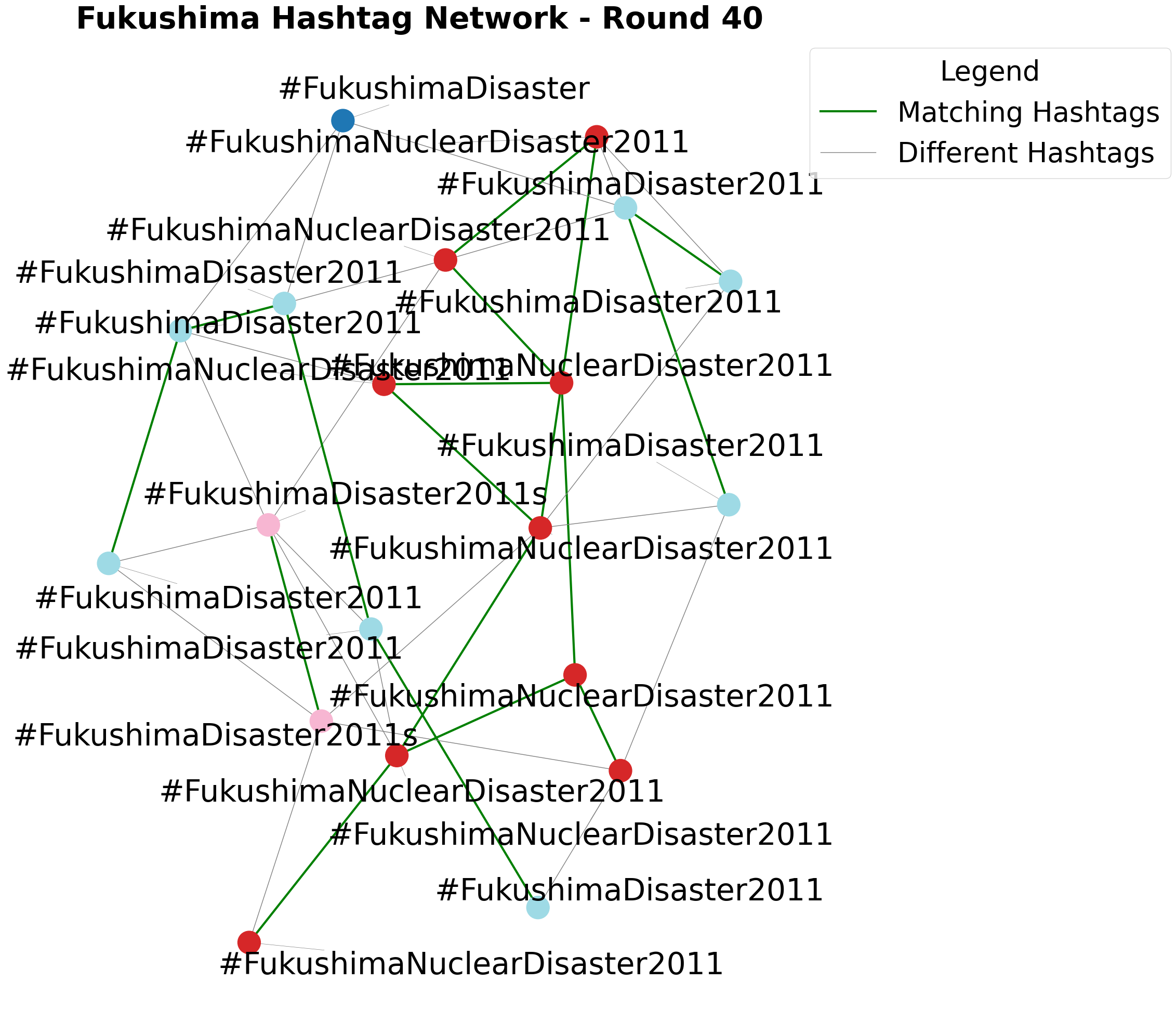}
        \caption{LLaMA Network — Round 40}
        \label{fig:network_structure_llama_round_40}
    \end{subfigure}
    \caption{Hashtags generated by the Llama network at the first and final trial.}
    \label{fig:network_structure_llama}
    \vspace{-1em}
\end{figure}

\subsection{Network Structure}

In both simulation studies, interactions between agents were structured using a Watts-Strogatz small-world network \cite{watts_strogatz}, which is characterized by high clustering and short average path lengths. This topology emulates real-world social networks by supporting both local coordination and global information diffusion, making it well-suited for modeling discourse convergence dynamics. At each round of interaction, agents are randomly assigned a new neighbor based on this network structure.

\section{Case Study 1: Fukushima Network Experiment}
\label{case_study_1}

In the first simulation, we compared LLM-generated data with human responses observed in a prior controlled experiment involving a hashtag matching game. In this task, both human participants and LLM agents were presented with a short narrative describing the 2011 Fukushima nuclear disaster. Participants were then instructed to generate hashtags that would align with those of their network neighbors, encouraging coordination within a structured social network. The human data was originally collected by Priniski et al., (2024) \cite{priniski2024online} and serves as a benchmark for evaluating the alignment and convergence patterns exhibited by LLMs. The full narrative used in the task is as follows:

\begin{quote}    
{\small
The Fukushima Nuclear Disaster was a 2011 nuclear accident at the Daiichi Nuclear Power Plant in Fukushima, Japan. The cause of the nuclear disaster was the Tōhoku earthquake on March 11, 2011, the most powerful earthquake ever recorded in Japan. The earthquake triggered a tsunami with waves up to 130 feet tall, with 45 foot tall waves causing direct damage to the nuclear power plant. The damage inflicted dramatic harm both locally and globally. 

The damage caused radioactive isotopes in reactor coolant to discharge into the sea, therefore Japanese authorities quickly implemented a 100-foot exclusion zone around the power plant. Large quantities of radioactive particles were found shortly after throughout the Pacific Ocean and reached the California coast.

The exclusion zone resulted in the displacement of approximately 156,000 people in years to follow. Independent commissions continue to recognize that affected residents are still struggling and facing grave concerns. Indeed, a WHO report predicts that infant girls exposed to the radiation are 70\% more likely to develop thyroid cancer.

The resulting energy shortage inspired media campaigns to encourage Japanese households and businesses to cut back on electrical usage, which led to the national movement Setsuden ("saving electricity"). The movement caused a dramatic decrease in the country's energy consumption during the crisis and later inspired the Japanese government to pass a battery of policies focused on reducing the energy consumption of large companies and households.
}
\end{quote}
\vspace{-2em}

\begin{figure*}[t]
    \centering
  
    \vspace{0.5cm}
    
        \centering
        \includegraphics[width=\linewidth]{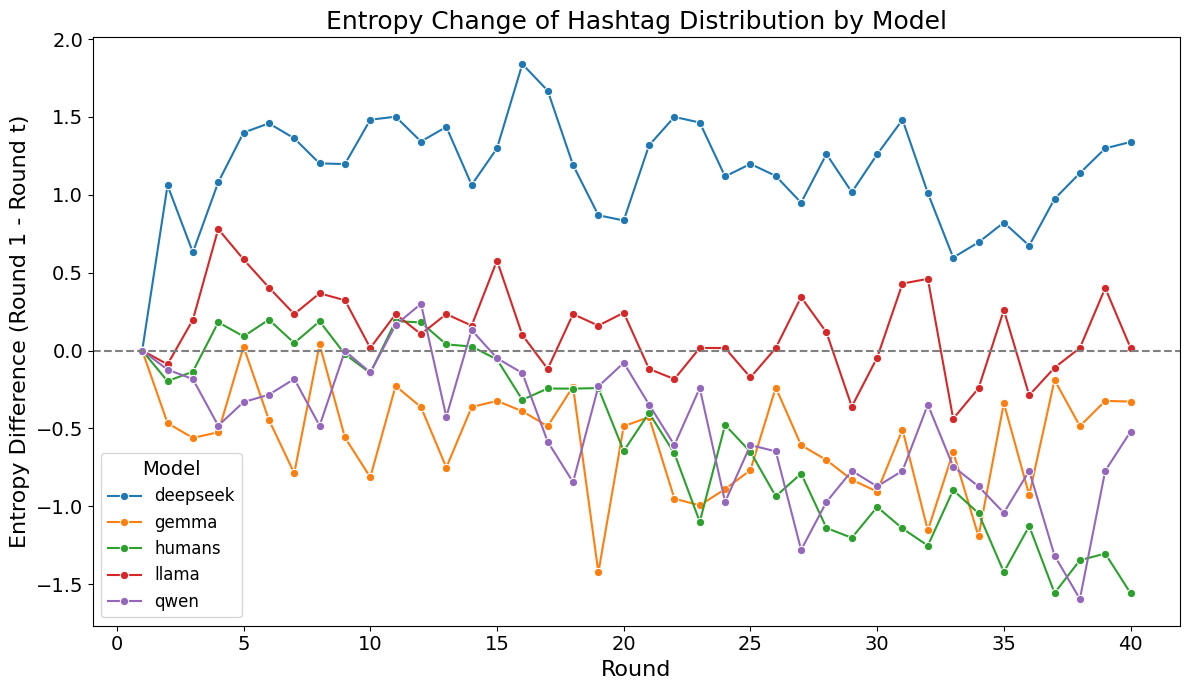}
        \caption{Change in entropy of each groups set of responses. Lower values indicate a large shift towards shared hashtag responses.}
        \label{fig:fuku_entropy}
        \vspace{-1em}
    \hfill
\end{figure*}

\begin{figure*}[t]
    \centering
    \begin{subfigure}[t]{0.45\textwidth}
        \includegraphics[width=\textwidth]{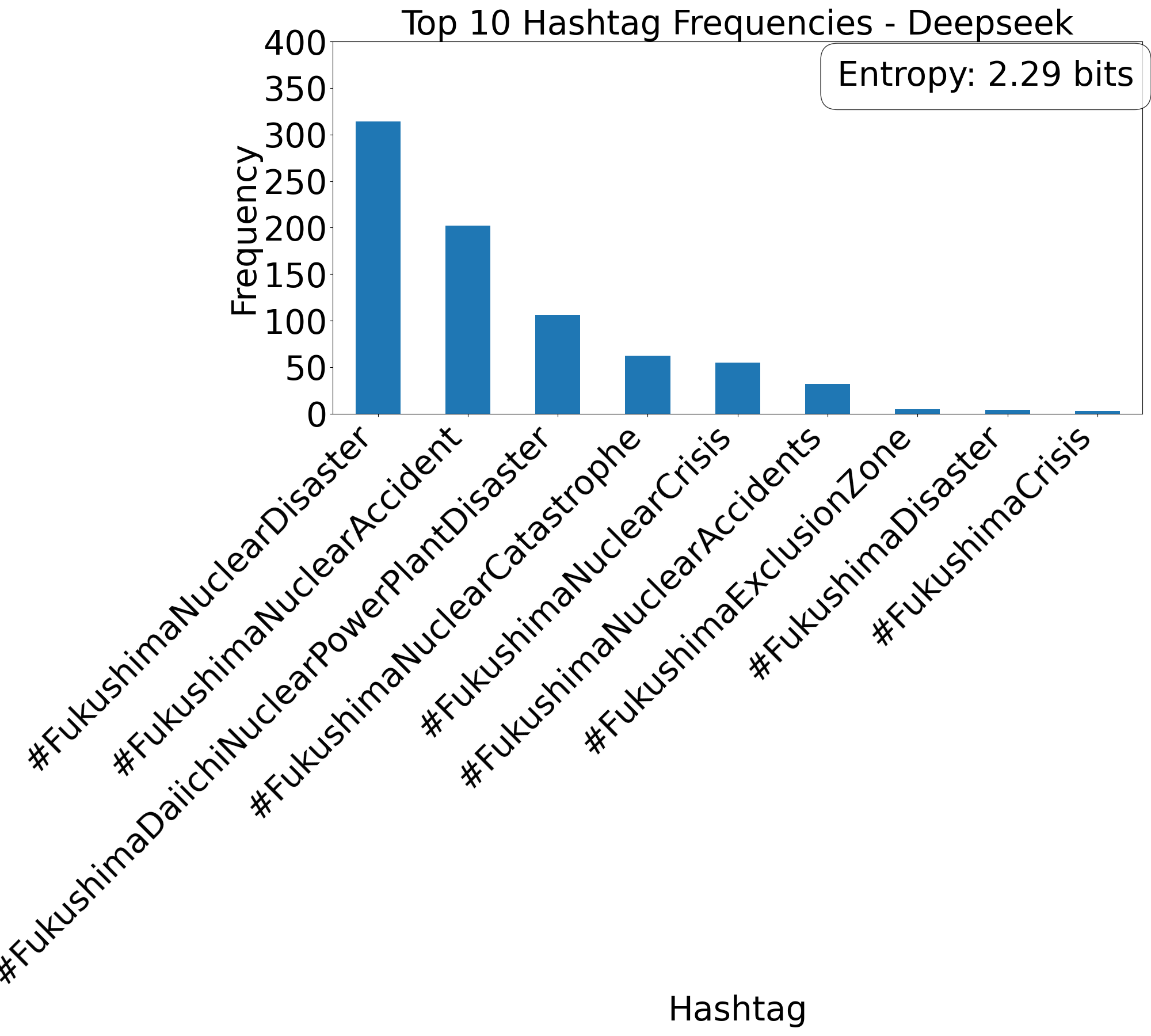}
        \caption{DeepSeek-R1}
    \end{subfigure}
    \hfill
    \begin{subfigure}[t]{0.45\textwidth}
        \includegraphics[width=\textwidth]{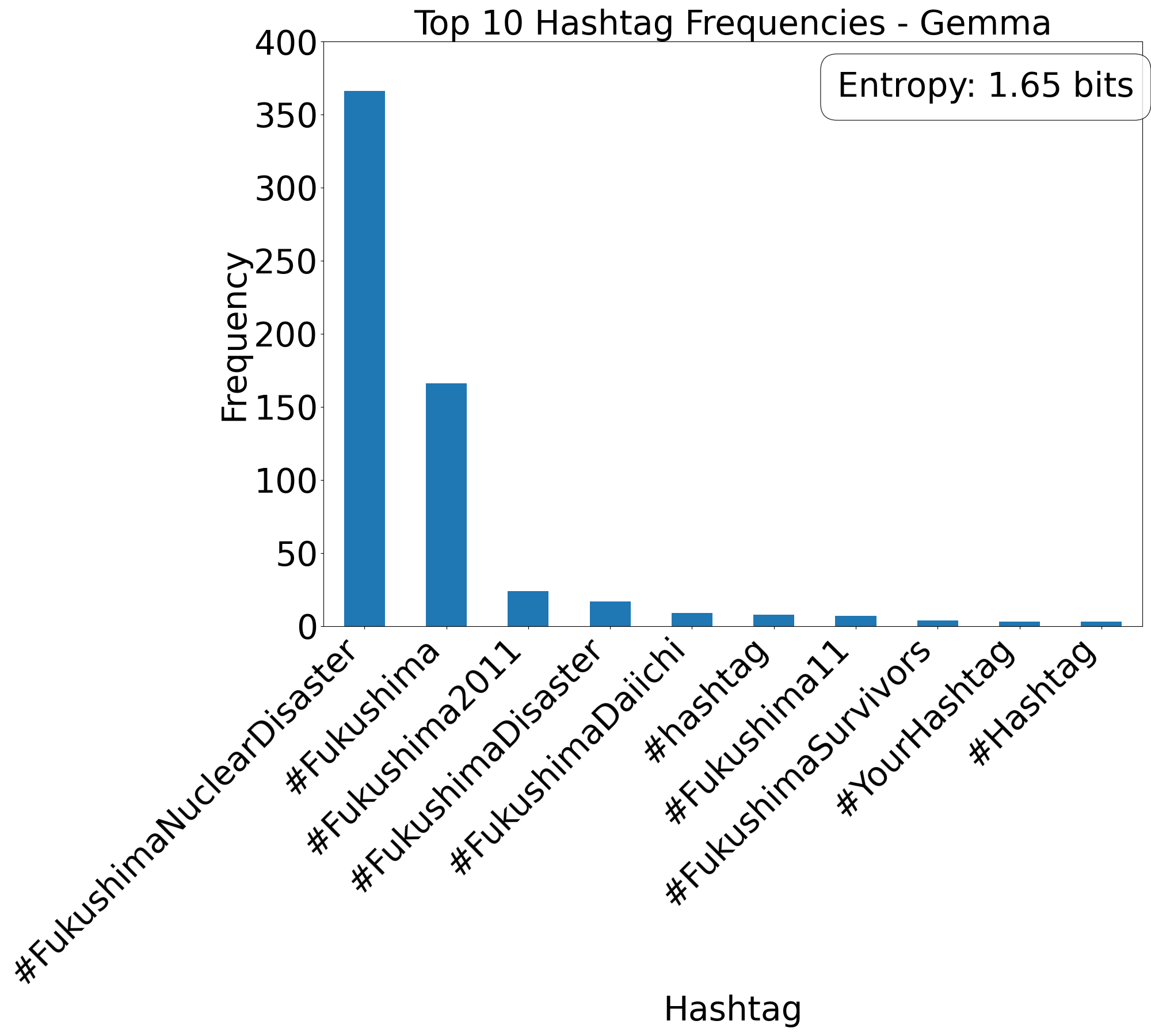}
        \caption{Gemma2}
    \end{subfigure}

    \vspace{1em}

    \begin{subfigure}[t]{0.45\textwidth}
        \includegraphics[width=\textwidth]{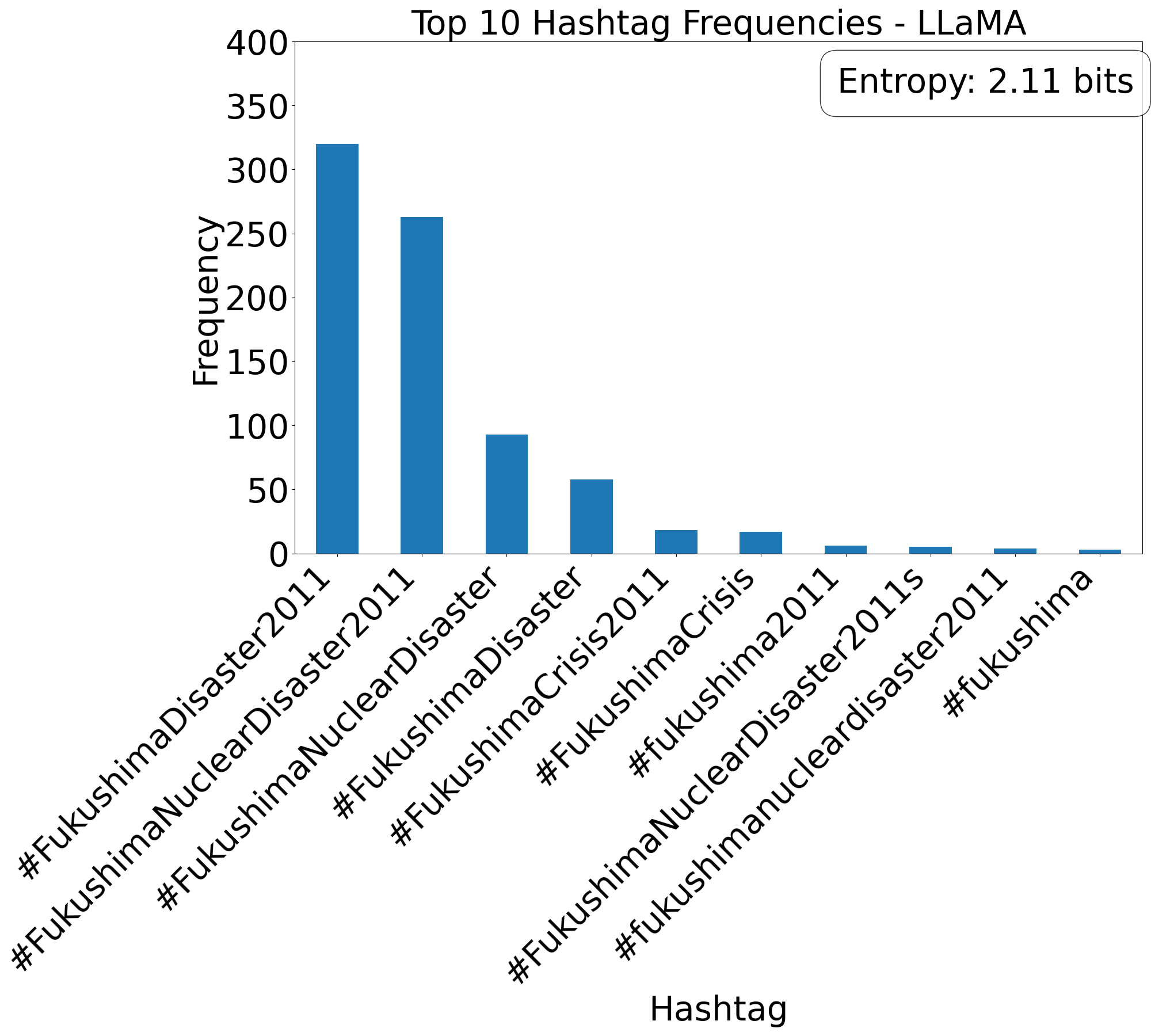}
        \caption{LLaMA-3}
    \end{subfigure}
    \hfill
    \begin{subfigure}[t]{0.45\textwidth}
        \includegraphics[width=\textwidth]{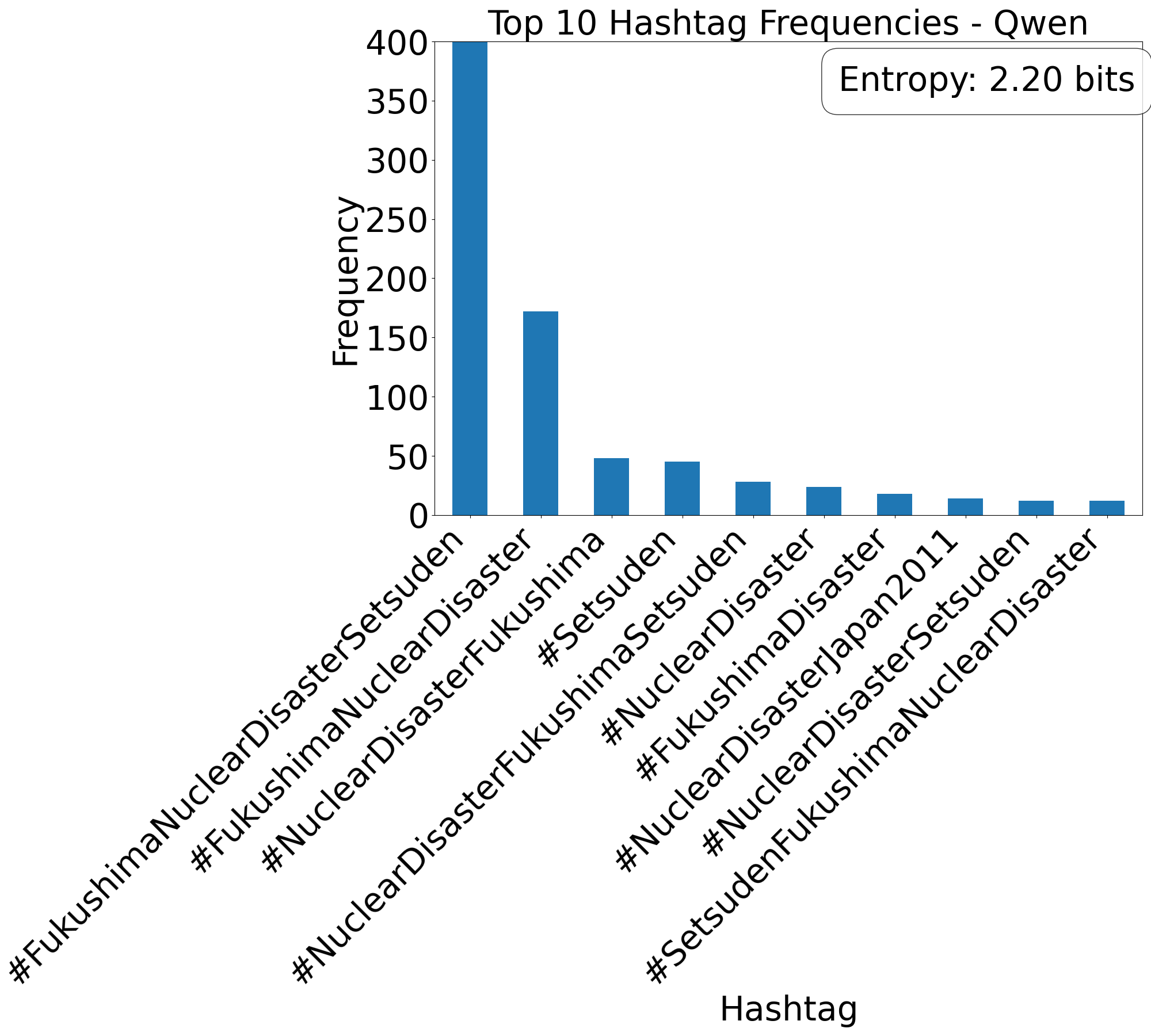}
       \caption{Qwen2}
    \end{subfigure}
  \caption{Rank-abundance Curves (RACs) of each LLMs ten most frequently generated hashtags in response to the Fukushima disaster narrative.}
    \label{fig:hashtag_freq_fukushima}
    \vspace{-1em}
\end{figure*}

\subsection{Results}
Figures~\ref{fig:network_structure_llama_round_1} and~\ref{fig:network_structure_llama_round_40} illustrate the variety of hashtags generated by agents at the beginning and end of networked interaction, respectively. In these visualizations, edges represent communication links between agents within each simulation trial. While these figures provide a snapshot of evolving alignment across the network in the Fukushima simulation, we applied three quantitative metrics to assess the onset and dynamics of shared hashtag usage over time:

\textbf{Group Entropy}: Drawing from metrics in population ecology \cite{avolio_comprehensive_2019,hallett_codyn_2016}), we computed an entropy score to quantify behavioral variability at the group level. Higher entropy values indicate greater diversity in group behavior (i.e., more variety in hashtags), while lower entropy values indicate a more coherent behavior (i.e., a more consistent set of hashtags). 

\textbf{Narrative Alignment}: To assess the extent to which agents' hashtags reflected elements of the event narrative, we employed an embedding-based narrative alignment procedure adapted from \cite{priniski2025effect}. The Fukushima narrative was segmented into discrete causal events, with each event’s textual description embedded using the SentenceTransformer model \texttt{all-MiniLM-L6-v2} \cite{all-minilm-l6-v2}. Hashtag responses were embedded similarly, and hashtags matched to the narrative event with the highest cosine similarity. This allowed us to track how semantically aligned agents’ outputs were with different parts of the underlying narrative.

\textbf{Perplexity}: To estimate how surprising the language generated by each model was, relative to naturalistic human hashtag use, we computed unigram perplexity scores. Specifically, we constructed a unigram language model based on the distribution of human-generated hashtags from the original experiment. We then measured the perplexity of model-generated hashtags with respect to this distribution. Perplexity quantifies the divergence between the observed outputs and expected human distribution: higher perplexity values indicate greater divergence and, by extension, less human-like linguistic behavior. This unigram model provides a simple yet interpretable proxy for linguistic plausibility, without introducing assumptions inherent to more complex language modeling approaches. By computing perplexity at each round of the simulation, we tracked how each model's hashtag usage evolved over time, either converging toward or diverging from human-like linguistic patterns as communication progressed.


  
    


\subsubsection{Onset of group coherence}

To assess the onset of group-level coherence (i.e., the degree to which individual nodes in a group produce the same behavior), we computed the entropy of each group’s hashtag distribution across simulation rounds. Lower entropy values indicate the emergence of a few dominant hashtags that account for most agent outputs, reflecting increased coherence. Conversely, higher entropy reflects a more even distribution across diverse hashtags, signaling less coordinated group behavior.

As illustrated in Figure \ref{fig:fuku_entropy}, entropy decreased steadily over time in human groups, suggesting a robust trend toward shared hashtag use. A similar but more gradual decline was observed in groups composed of Gemma and Qwen agents, indicating moderate convergence. In contrast, entropy remained relatively flat in groups using the LLaMA model and increased in DeepSeek-based groups, suggesting limited or even deteriorating coordination over time. These patterns indicate that Gemma and Qwen were more effective at adapting to social signals during communication, whereas LLaMA and DeepSeek struggled to integrate neighbor input in a way that fostered coherent group behavior.

\begin{figure}[t]
    \centering
    \includegraphics[width=.85\linewidth]{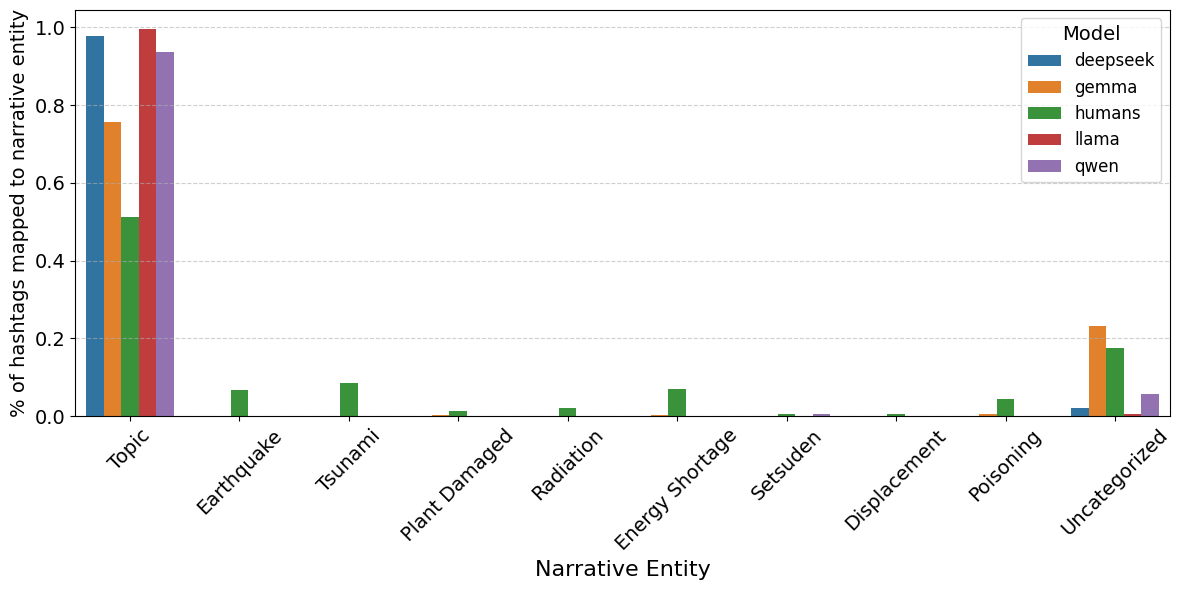}
    \caption{Number of hashtags most similar to each narrative entity. While humans produce hashtags that align with both semantic topics (e.g., Fukushima Disaster) and cause and effect relationships, each of the LLMs nearly exclusively produced hashtags about topics. }
    \label{fig:narrative-align-fuku}
    \vspace{-1em}
\end{figure}

\subsubsection{Perplexity} 
\begin{figure}[t]
    \centering
    \begin{subfigure}[t]{0.48\textwidth}
        \centering
        \includegraphics[width=\textwidth]{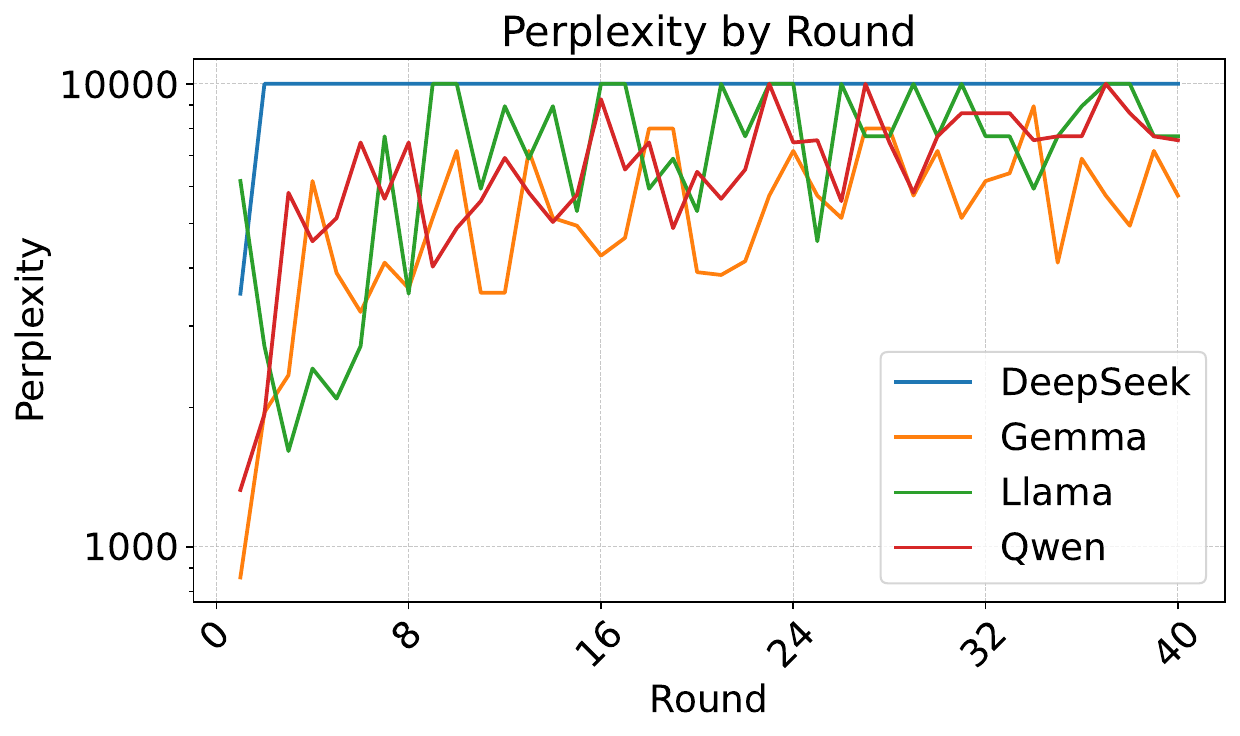}
        \caption{Fukushima}
        \label{fig:perplexity-fuku}
    \end{subfigure}
    \hfill
    \begin{subfigure}[t]{0.48\textwidth}
        \centering
        \includegraphics[width=\textwidth]{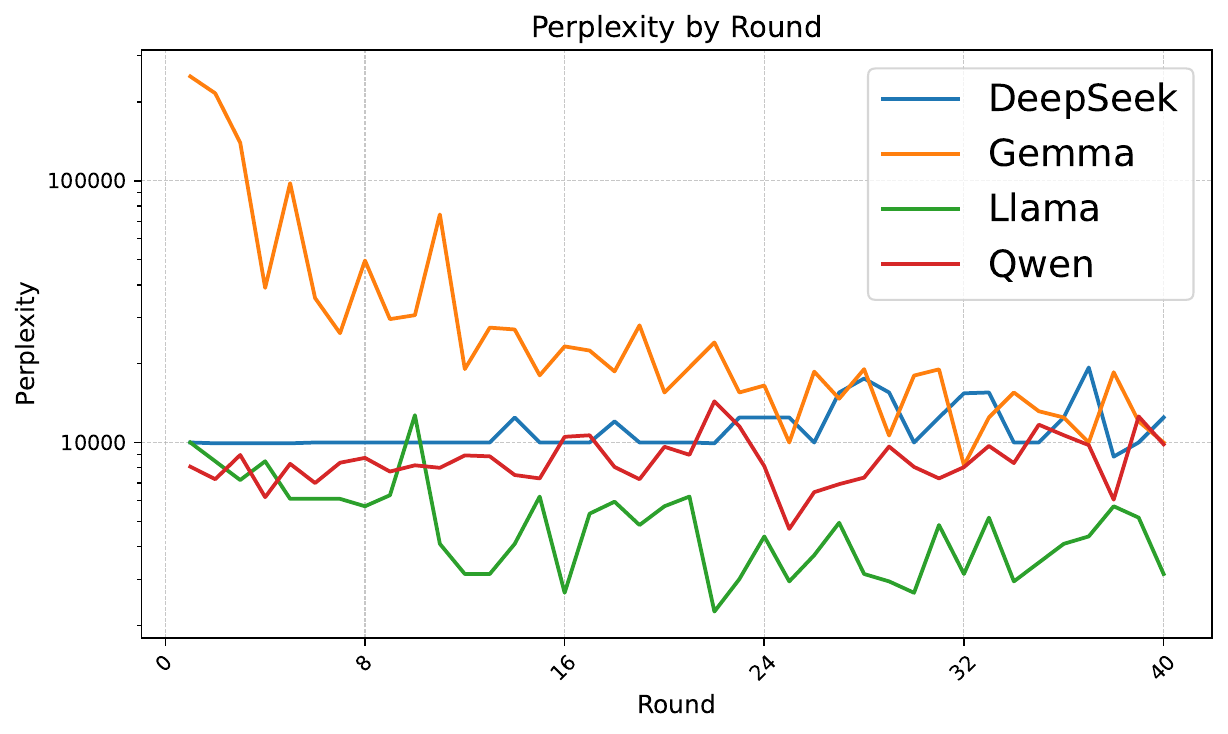}
        \caption{Philippines Election} \label{fig:perplexity-phil}
    \end{subfigure}
    \caption{Perplexity by round for each setting.}
    \label{fig:perplexity}
    \vspace{-1em}
\end{figure}

In the Fukushima experiment,the human-generated hashtags from the original network trials served as the basis for constructing the unigram reference distribution. As shown in Figure~\ref{fig:perplexity-fuku}, perplexity scores increased across rounds for most models, indicating that their outputs became progressively less aligned with human-generated language over time. This upward trend suggests that iterative exposure to peer behavior did not aid the models in converging toward human-like language patterns. On the contrary, continued interaction often led models to generate hashtags that diverged further from human norms.

\subsubsection{Narrative alignment of hashtags}

Beyond analyzing the distribution of discrete hashtag values, we conducted an embedding-based analysis to assess how hashtags aligned with key narrative entities from the Fukushima nuclear disaster. As shown in Figure \ref{fig:narrative-align-fuku}, human participants generated hashtags that captured both semantic topics (e.g., Fukushima Disaster) and cause and effect relationships (e.g., Earthquake, Displacement). In contrast, each of the LLMs predominantly produced topic-oriented hashtags.

To understand the semantic diversity of hashtag production across groups, we analyzed the Rank Abundance Curves (RACs) of each model's most frequent hashtags. Figure~\ref{fig:hashtag_freq_fukushima} presents the frequencies of the ten most frequently generated hashtags by each model. As shown in Figure \ref{fig:hashtag_freq_fukushima}(d), the Qwen language model produced hashtags that merged topic-level hashtags (e.g., Fukushima Nuclear Disaster) with the downstream effect event (Setsuden). Merging the strings for these separate entities likely reduced coordinate rates for the model. These findings suggest that LLMs exhibit a bias toward producing topic-oriented hashtags and may require structured prompting to elicit outputs that align with causal relationships expressed in the narrative. Additionally, prompt designs should encourage parsimonious hashtag construction—favoring shorter, more specific strings that reduce ambiguity and improve coordination within agent networks.

\section{Study 2: Philippine 2022 Presidential Election}

In the second simulation, we examine political discourse coordination by simulating LLM-based agent behavior during the 2022 Philippine Presidential Election. This setting allows us to investigate how generative agents perform in a contested, diverse political context, where rhetorical positioning and group alignment are key dynamics. The following event description was used to provide all agents with a shared narrative context:


\begin{quote}

{\small
    In 2022, the Philippines held a national election to choose its next president, following six years under President Rodrigo Duterte. The election featured candidates each offering different perspectives on the country’s direction in areas such as economic recovery, foreign policy, governance, and national identity. Voters were presented with a diverse range of platforms, leadership styles, and policy priorities. Candidates for President: Ferdinand “Bongbong” Marcos Jr. – Partido Federal ng Pilipinas (PFP), Leni Robredo – Independent (Liberal Party member), Manny Pacquiao – PROMDI, Isko Moreno – Aksyon Demokratiko, Panfilo “Ping” Lacson – Independent (formerly Partido Reporma), Leody de Guzman – Partido Lakas ng Masa (PLM), Faisal Mangondato – Katipunan ng Kamalayang Kayumanggi (KKK), Jose Montemayor Jr. – Democratic Party of the Philippines (DPP), Norberto Gonzales – Partido Demokratiko Sosyalista ng Pilipinas (PDSP), Ernesto Abella – Independent
}
\end{quote}
\vspace{-2em}

\begin{figure*}[t]
    \centering
        \centering
        \includegraphics[width=.75\linewidth]{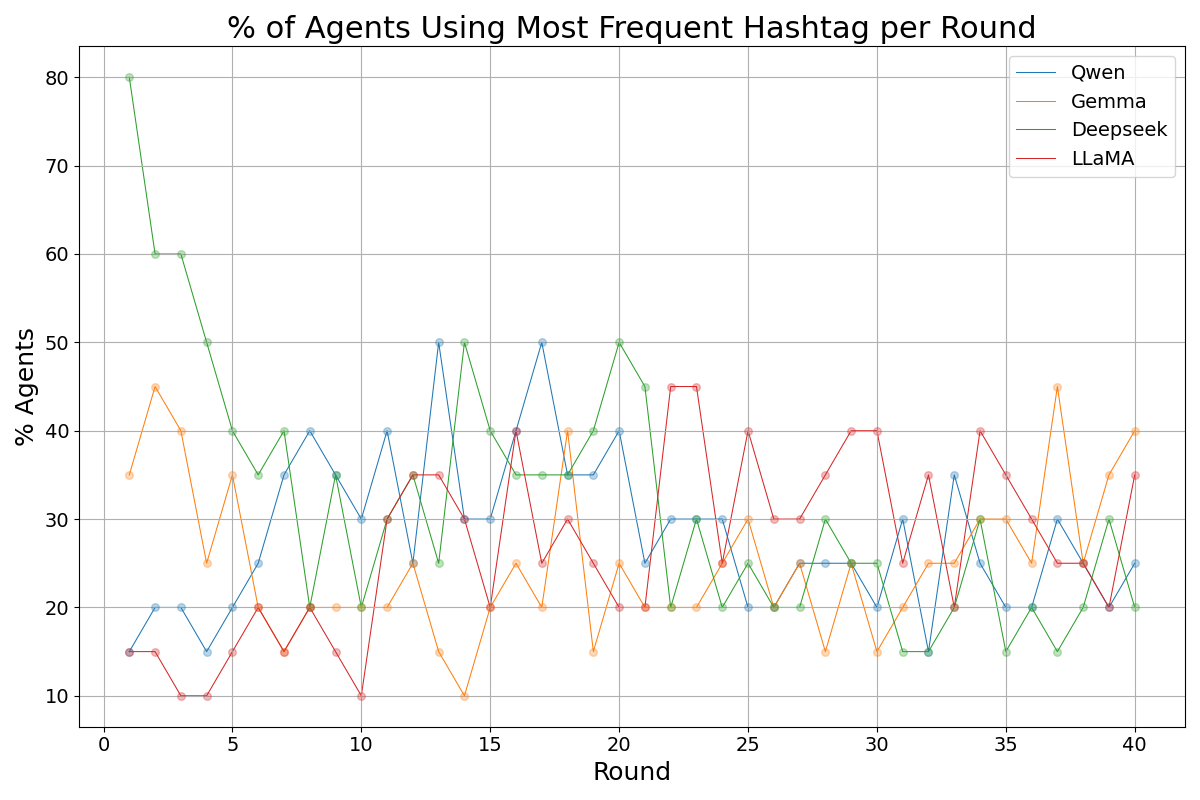}
        \caption{Percentage of agents producing the dominant hashtag in each round for the Philippine election simulation.}
        \label{fig:election_hashtag_alignment}
    \vspace{-1em}
\end{figure*}

\begin{figure}[t]
    \centering
    \begin{subfigure}[t]{0.48\textwidth}
        \centering
        \includegraphics[width=\textwidth]{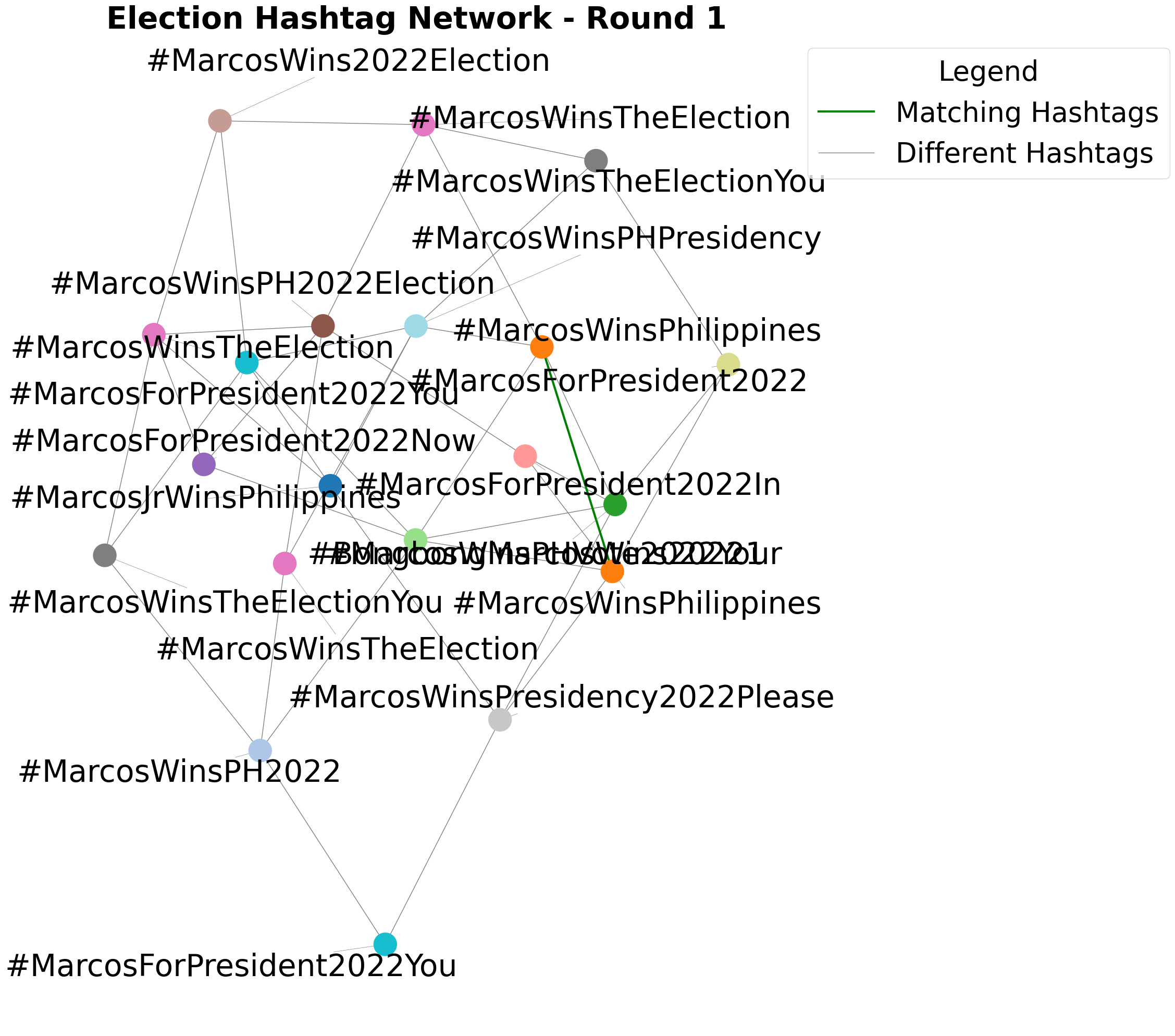}
        \caption{LLaMA Network — Round 1}
        \label{fig:network_structure_llama_round_1_election}
    \end{subfigure}
    \hfill
    \begin{subfigure}[t]{0.48\textwidth}
        \centering
        \includegraphics[width=\textwidth]{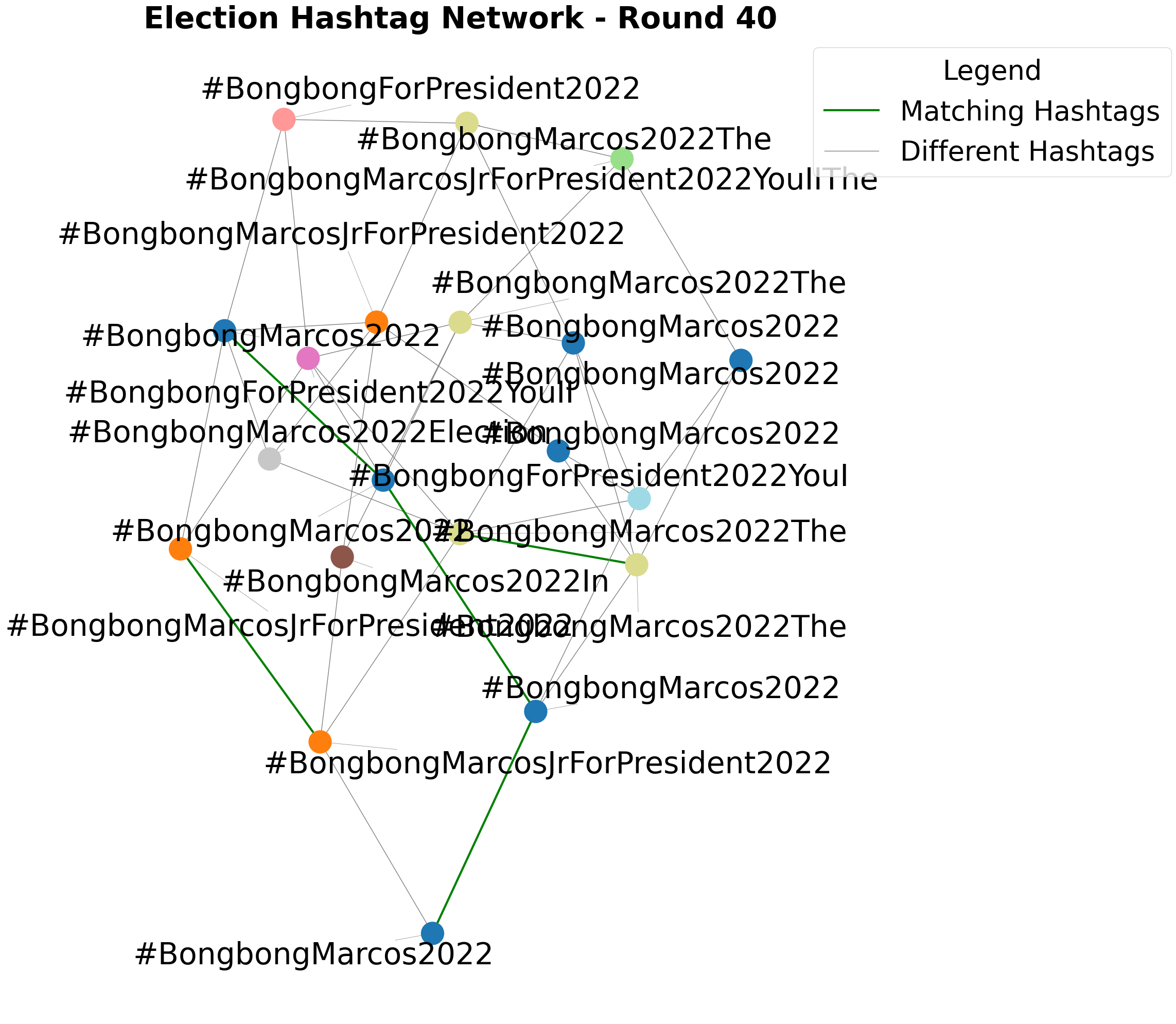}
        \caption{LLaMA Network — Round 40}
        \label{fig:network_structure_llama_round_40_election}
    \end{subfigure}
    \caption{Network structure evolution over time in the Phillipine election simulation for the LLaMA model.}
    \label{fig:network_structure_llama_election}
    \vspace{-2em}
\end{figure}

\subsection{Results}
 Figures~\ref{fig:network_structure_llama_round_1_election}  and~\ref{fig:network_structure_llama_round_40_election} illustrate the variety of hashtags agents generated at the beginning and end of networked interaction, and edges show the linked nodes are those respective trials.  While this figure provides a snapshot on the hashtags being generated, we applied the same quantitative measures in the previous study to better assess hashtag convergence dynamics based on communication over Philippine election narrative. 
 

\subsubsection{Onset of Group Coherence}

As in the Fukushima study, we evaluated group-level convergence by computing the proportion of agents in each group producing the normative (i.e., most frequently generated) hashtag. Higher proportions indicate stronger convergence toward a shared framing of the event. As shown in Figure~\ref{fig:election_hashtag_alignment}, none of the models demonstrated strong coordination, with only 10–50\% of agents producing the same response. Notable, Notably, DeepSeek initially exhibited high convergence, driven by frequent generation of \#PhillipinesElection2022, but this coherence steadily declined over rounds. This suggests that ongoing interaction and social exposure increased uncertainty rather than reinforcing group consensus. This suggests that more structured prompts are necessary to help models efficiently integrate background knowledge about an event (as encoded by the group's shared focal narrative). 

\begin{figure*}[t!]
    \centering
    \begin{subfigure}[t]{0.45\textwidth}
        \includegraphics[width=\textwidth]{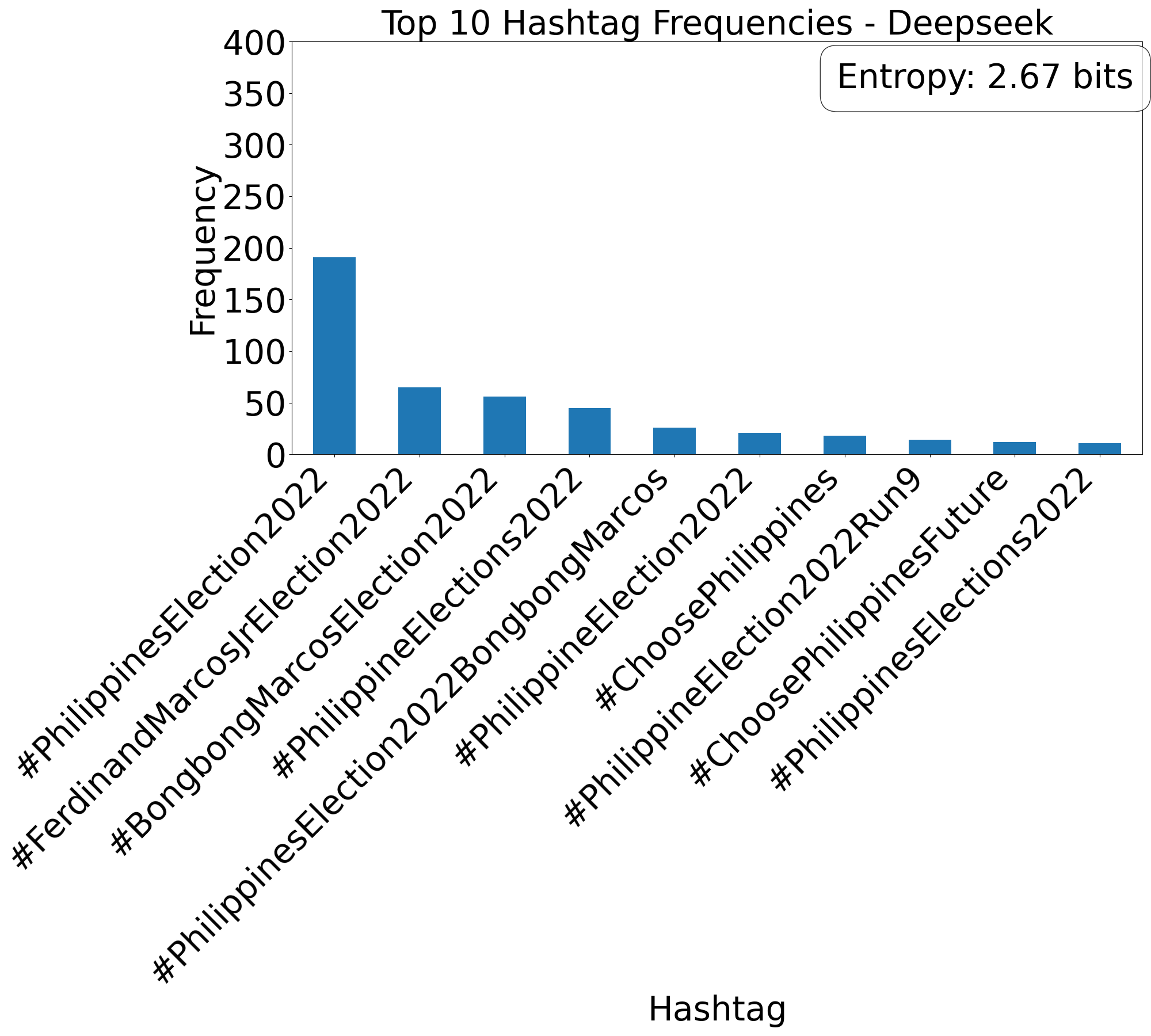}
        \caption{DeepSeek-R1}
    \end{subfigure}
    \hfill
    \begin{subfigure}[t]{0.45\textwidth}
        \includegraphics[width=\textwidth]{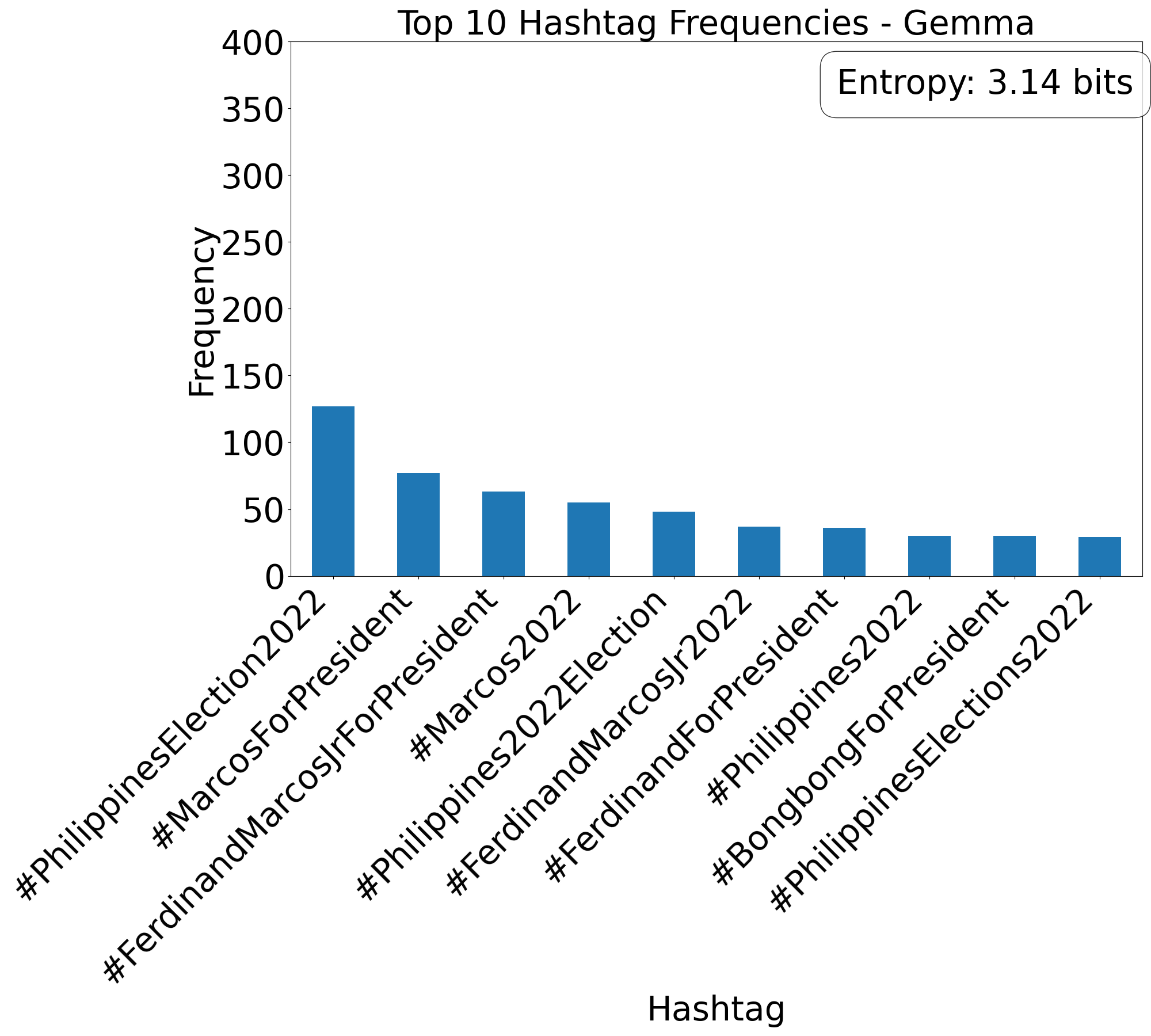}
        \caption{Gemma2}
    \end{subfigure}

    \vspace{1em}

    \begin{subfigure}[t]{0.45\textwidth}
        \includegraphics[width=\textwidth]{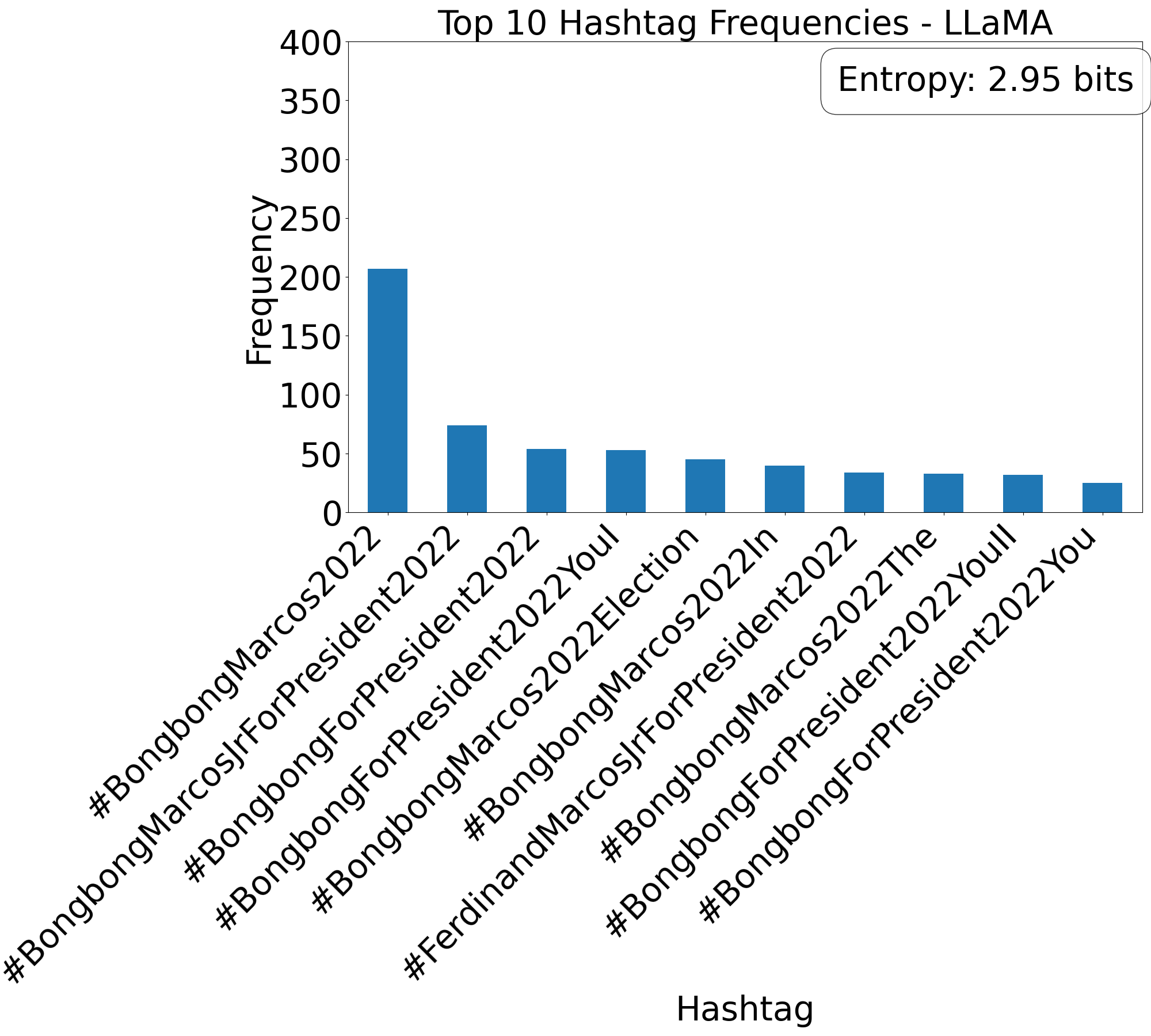}
        \caption{LLaMA-3}
    \end{subfigure}
    \hfill
    \begin{subfigure}[t]{0.45\textwidth}
        \includegraphics[width=\textwidth]{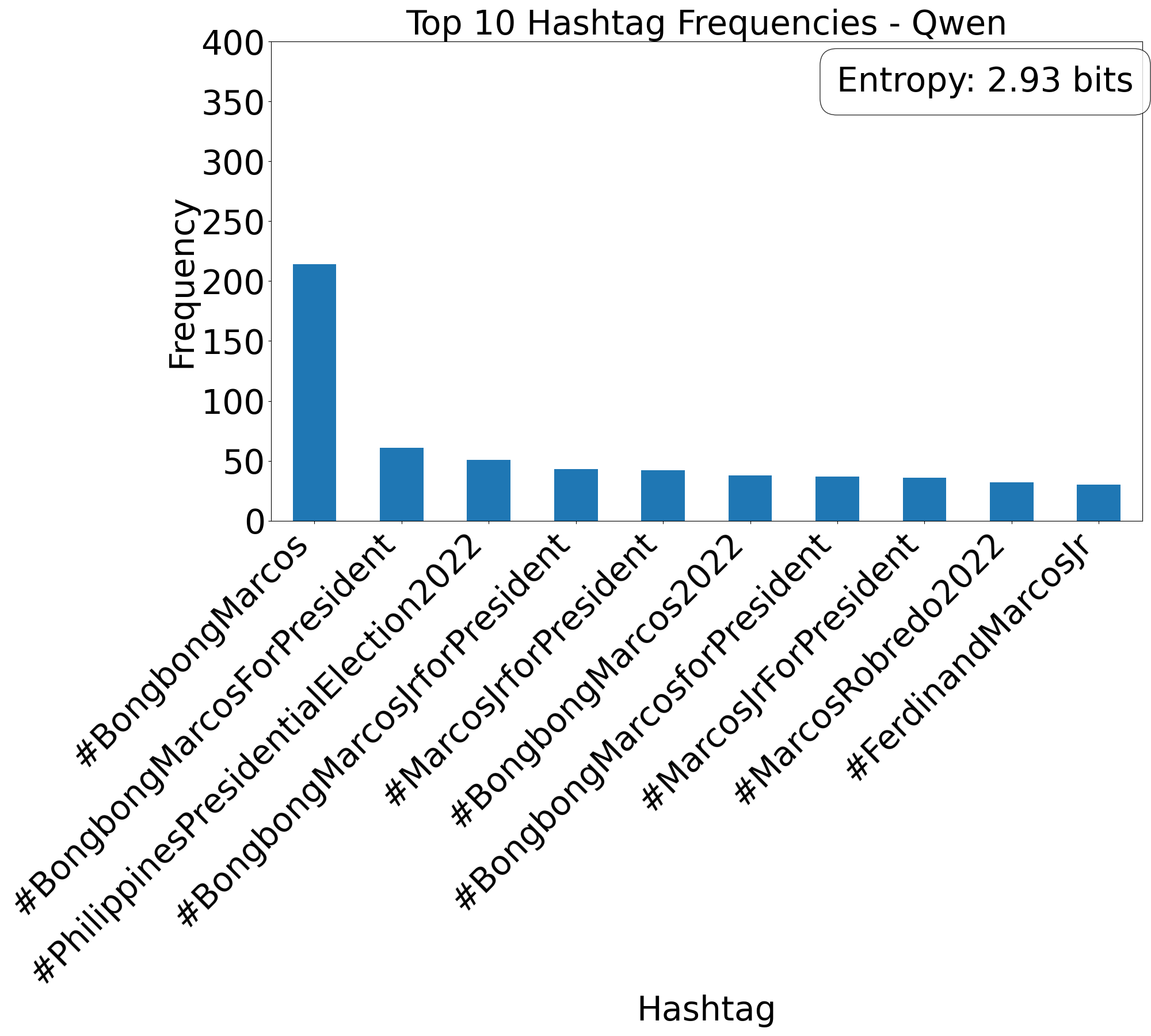}
        \caption{Qwen2}
    \end{subfigure}

    \caption{Top 10 hashtag frequency distributions for the Philippine election simulation across four LLMs.}
    \label{fig:hashtag_freq_election}
    \vspace{-2em}
\end{figure*}

To complement this measure, we also computed the entropy of the full hashtag distribution produced by each model across all rounds. Figure~\ref{fig:hashtag_freq_election} presents the Rank Abundance Curves (RACs) for the ten most frequently generated hashtags, with each subplot displaying the entropy of the corresponding model's full hashtag distribution. Entropy values across models hovered around $3$, and the top hashtags (e.g., \#BongbongMarcos2022 from Deepseek and \#BongbongMarcos from Gemma2) each occurred approximately 200 times. These entropy values were notably higher than those observed in the Fukushima simulation, suggesting that the election narrative imposed weaker constraints on model behavior and permitted greater lexical variability. Further inspection of the content of top hashtags reveals that models disproportionately focused on the first candidate listed in the prompt. This pattern suggests that LLMs defaulted to salient or primacy-biased entities when left unguided. To improve coordination and fairness in simulated political discourse, future prompting strategies should consider randomizing candidate order or explicitly instructing agents to sample across available options.

\vspace{-2em}
\subsubsection{Perplexity}

To assess how closely model-generated hashtags aligned with naturally occurring online discourse, we constructed a unigram language model from a large corpus of Twitter hashtags collected between January and May 2022. We then computed the perplexity of each model’s outputs at every round of the simulation to track how linguistic alignment evolved over time.

Unlike in the Fukushima setting, the perplexity scores generally decreased with time for most models (Figure~\ref{fig:perplexity-phil}). Gemma and LLaMA showed the most consistent improvement, suggesting that repeated exposure to peer responses led them to generate increasingly human-like outputs. This contrast with the Fukushima results indicates that models may perform better in simulating discourse for domains that closely match their pre-training data (e.g., social networks). These findings underscore the value of iterative interaction in steering model behavior toward naturalistic language use.

\section{Conclusion}
\vspace{-1em}
Our simulations show that LLM-based agents can exhibit partial convergence in networked coordination tasks, but often fall short of replicating the onset of group coherence observed in human groups. Although models like Gemma and Qwen2 showed a decrease in entropy over time when coordianting hashtags over the Fukushima nuclear disaster,  other models like DeepSeek and LLaMA-3 either stagnated or diverged, indicating weak social alignment. Furthermore, convergence was weak when generating hashtags about the Philippines election narrative, with the majority of the hashtags centering around the first listed candidate, which suggests that additional prompting may be necessary for agents to reach hashtag consensus in addition to generating ideologically positioned hashtags (i.e., choosing one candidate over another).  

Aligning the hashtags to the semantic content of the focal narrative in the Fukushima nuclear disaster study revealed a strong bias across all LLMs toward topic-level hashtags, whereas humans captured both topical and causal dimensions. This suggests that LLMs struggle to model deeper narrative structures including causal and effect relations across narrative entities. Additionally, model perplexity trends diverged across the two studies. Fukushima simulations showed an increasing divergence from human patterns, while in the Philippines scenario, convergence improved over time, possibly because the language used in that setting more closely resembles the social media content included in the models’ pretraining data. All together, these results highlight the need for better prompting strategies and model alignment when simulating group communication with LLMs.
\vspace{-1em}

\section*{Acknowledgments}
This research was supported, in part, by AFOSR MURI grant \#FA9550-22-1-0380, and by a grant from the U.S. Government \#65000025C0014.

\begingroup
\let\clearpage\relax
\bibliographystyle{ACM-Reference-Format}
\bibliography{references}
\appendix
\endgroup

\end{document}